\documentclass[%
preprint,
amsmath,
amssymb,
 aps,
prl,
]{revtex4-2}

\usepackage{setspace}

\usepackage[LGR,T1]{fontenc} 
\usepackage[utf8]{inputenc} %
\usepackage{substitutefont} 
\usepackage[polutonikogreek,english]{babel}
\usepackage[largesc]{newtxtext} %
\usepackage[varqu,varl]{zi4}
\usepackage{cabin}
\usepackage[vvarbb,upint,slantedGreek,smallerops]{newtxmath}
\substitutefont{LGR}{\rmdefault}{Tempora} 

\usepackage{textcomp}

\usepackage[margin=1.5in]{geometry}
\usepackage{hyperref}
\usepackage[hyphens]{xurl}
\usepackage{amsmath}
\usepackage{amsfonts}
\usepackage{amssymb}

\usepackage[pdftex]{graphicx}
\usepackage{lastpage}
\usepackage{slashed}

\usepackage{xcolor}

\usepackage{xfrac}

\usepackage{comment}
\usepackage[yyyymmdd,hhmmss]{datetime} 

\usepackage{lastpage}
\usepackage[running,mathlines]{lineno}
\usepackage{isomath}
\usepackage{siunitx}
\usepackage{upgreek}

\usepackage{marginnote}

\usepackage{amsthm}

\newtheorem{theorem}{Theorem}

\newtheorem{lemma}{Lemma}

\newtheorem*{law*}{Law}

\newtheorem{conjecture}{Conjecture}

\newtheorem{proposition}{Proposition}

\theoremstyle{definition}

\usepackage{verbatim}

\usepackage{xspace}
\usepackage{braket}
\usepackage{mathtools}
\mathtoolsset{showonlyrefs=true}

\usepackage{siunitx}
\usepackage{physics}

\newcommand{\pr}{\ensuremath{\mathrm{P}}}

\newcommand{\piu}%
{\textrm{\greektext p}}
\newcommand{\eu}%
{\ensuremath{\mathrm{e}}}
\newcommand{\iu}%
{\ensuremath{\mathrm{i}}}

\providecommand{\newoperator}[3]{%
\newcommand*{#1}{\mathop{#2}#3}}

\newcommand{\tran}%
{\textsf{T}}
\newcommand{\herm}%
{\textsf{H}}
\newcommand{\deltau}%
{\textrm{\greektext d}}
\newcommand{\Deltau}%
{\textrm{\greektext D}}

\makeatletter
\providecommand*{\diff}%
{\@ifnextchar^{\DIfF}{\DIfF^{}}}
\def\DIfF^#1{%
\mathop{\mathrm{\mathstrut d}}
\nolimits^{#1}\gobblespace}
\def\gobblespace{%
\futurelet\diffarg\opspace}
\def\opspace{%
\let\DiffSpace\!%
\ifx\diffarg(%
\let\DiffSpace\relax
\else
\ifx\diffarg[%
\let\DiffSpace\relax
\else
\ifx\diffarg\{%
\let\DiffSpace\relax
\fi\fi\fi\DiffSpace}

\newoperator{\loglike}%
{\mathrm{loglike}}{\nolimits}

\newoperator{\notloglike}%
{\mathrm{notloglike}}{\limits}

\makeatother

\renewcommand{\piu}{\uppi}
\renewcommand{\deltau}{\updelta}

\let\originalpartial\partial
\let\partial\relax
\newrobustcmd*{\partial}{\text{\rotatebox[origin=t]{10}{\scalebox{0.95}[1]{\ensuremath{\originalpartial}}}}\hspace{-0.05em}}

\usepackage{graphicx}
\usepackage{dcolumn}

\usepackage[normalem]{ulem}

\usepackage{bm}

\newcommand{\vN}{von Neumann\xspace}

\newcommand{\kernBeforeIntegral}{\ensuremath{\kern-0.17em}\xspace}

\newcommand{\eqdefA}{=\xspace}

\newcommand{\entropyS}{\ensuremath{\mathrm{S}\xspace}}

\newcommand{\QFT}{\ensuremath{\text{QFT}}\xspace}

\usepackage[mathscr]{euscript}

\usepackage[en-US]{datetime2}

\usepackage{nicefrac}

\usepackage[inline]{enumitem}

\DeclareMathOperator{\Mod}{mod}

\usepackage{changepage}
\begin{document}

\thispagestyle{empty}
\newpage

\title{Spin-Entropy}
\author{{Davi Geiger} and {Zvi M.\ Kedem}}
\affiliation{ Courant Institute of Mathematical Sciences\\
 New York University, New York, New York 10012}

\begin{abstract}

In classical physics, entropy quantifies the randomness of large systems, where the complete specification of the state, though possible in theory, is impossible in practice. In quantum physics, despite its inherently probabilistic nature, the concept of entropy has been elusive.  The von Neumann entropy, currently adopted in quantum information and computing, models only the randomness associated with unknown specifications of a state and is zero for pure quantum states, and thus cannot quantify the inherent randomness of its observables. Our goal is to provide such quantification.

This paper focuses on the quantification of the randomness associated with observed spin values of a pure quantum, given an axis $z$. To this end, we define a spin-entropy whose minimum is $\ln 2\piu$, reflecting the uncertainty principle for the spin observables. We also extend the concept to quantum mixed states. 

The  spin-entropy attains local minima  for  entangled Bell states,  and local maxima for disentangled states. The  spin-entropy  may be useful for analyzing physical phenomena and developing robust quantum computational processes.

\end{abstract}

\maketitle

\pagebreak

\tableofcontents

\pagebreak

\section{Introduction}

The quantification of randomness associated with a quantum state is a subject that bears analogies with the classical state. In classical physics, entropy was introduced to handle large ensembles of particles, where the complete specification of the state is  not practical. However,  it has been an elusive concept in quantum physics.  Von Neumann entropy \cite{von2018mathematical}  quantifies the lack of knowledge an observer has about the quantum state, i.e., it quantifies the  randomness in  specifying the degrees of freedom (DOFs) of a quantum state.  However, it  does not  quantify  the intrinsic randomness of the observables associated with a  quantum state whose DOFs are fully known. Thus, for all single particles and for all entangled particles   \vN entropy is zero,  even though only probabilities  are knowable about the set of observables.    
The Stern–Gerlach experiment~\cite{gerlach1922experimentelle} illustrates the quantum scenario for spin, where  spin $\nicefrac{1}{2}$ particles with a  $z$ up direction state are prepared and, despite having zero von Neuman entropy, randomness on the spin values along any  direction perpendicular to $z$  is observed. Clearly, this uncertainty is due to the intrinsic randomness of the observables given the prepared  quantum state,  the fully specified  $z$ up state.

Our focus is  spin systems.  We propose a definition of spin-entropy in order to quantify the  randomness of the observables once the DOFs of a quantum spin state have been specified.    We adopt the geometric quantization method of the spin to create a spin phase space,  in which the randomness of the state can be quantified. We show that this entropy has the minimum value of $\ln 2\piu$, due to the  uncertainty principle for spin. We extend the definition to mixed states and to quantum field theory where the number of particles is also an observable. 

In this new quantification of randomness, we explore spin entanglement and disentanglement of two  particles.  Entanglement states tend to disentangle  in nature, posing  a challenge to build quantum computing processes, and thus maintaining entanglement is  a  topic of much interest  in quantum information and quantum computing. A better understanding of temporal evolution of such states will influence the choice of quantum processes  for constructing robust quantum algorithms and computers. For the case of two particles, the more entangled the states are the lower is the entropy.  For  three fermions of spin $\nicefrac{1}{2}$ we suggest that maximum entangled states should be defined by the entropy value. The lower the entropy of an entangled state,  the greater the entanglement.

\subsection{Previous Work}
Wehrl entropy \cite{wehrl1978general} was introduced to  approximate a classical entropy from a quantum state. Lieb studied coherent spin states to evaluate Wehrl entropy for spin states \cite{lieb:91,lieb2014proof}. As in the case of spatial coordinates, spin coherent states constitute an overcomplete set of states. For spin \nicefrac{1}{2}, all spin states up to a global phase are coherent states. For spin $1$ a large set of states are coherent states, including  two of the eigenstates of the spin operator $S_z$.  Wehrl entropy is minimized for all such coherent states\cite{lieb2014proof, Schupp.99}.   

When examining entangled and disentangled states,  Wehrl entropy is minimized for disentangled states. However,  reflecting on the randomness of the observables, and due to the entangleness, the entropy of entangled states should be lower  than that of  disentangled ones.  
We argue that such general overcomplete basis and the arbitrariness of the choice of coherent states to define the probability distribution, prevents Wehrl entropy from accurately quantifying  the randomness associated with the spin state observables.  We argue that  a proper probability distribution satisfying all Kolmogorov axioms of a probability is required to quantify the randomness of the observables.

For the case of two particles, the behaviors of \vN entropy and Wehrl entropy of entanglement have been studied by tracing out the states of one particle  to obtain a mixture of states for the other particle.  A salient difference between our proposed entropy and \vN entropy and Wehrl entropy is exhibited  during the temporal evolution of an entangled state to a disentangled one.  While our proposed entropy increases the more  the state disentangle, the \vN entropy and Wehrl entropy decreases the more the states disentangle.

\section{Spin Phase Space}
\label{sec:spin-geometry}
We first briefly review geometric aspects of spin, including the geometric quantization method that leads to the ppin phase space.

The spin matrix associated with a particle can be specified, e.g., \cite{cohen2019quantum}, as
\begin{align}
    \vec{S} & =S_x\hat x +S_y\hat y+ S_z\hat z \quad \text{and} \quad  S^2  = S_x^2+S_y^2+S_z^2\,,
    \\
[S_a,S_b] &= \iu \hbar S_c\,, \quad \text{where $a,b,c$ is a cyclic permutation of $x, y,z$}\,, \, \text{and}
\\
[S^2,S_a]&=0\,, \quad \text{for $ a=x,y,z$}\,.
\label{eq:spin-commutation}
\end{align}
The spin value of a particle is a Casimir  invariant  but it is not possible to know the values of the spin  projections in all directions in  three dimensions.  Knowing the value of the spin along the $z$ direction  implies that the $x$  or the $y$ direction spin values are unknowable. This uncertainty  reflects the close relation between spin matrices, their unitary transformations,  and the rotation group $\text{SO}(3)$. For spin~$\nicefrac{1}{2}$ particles, any spin state is reachable from any other spin state via a  $2\times 2$ unitary transformation, which is a local isomorphism (and a global homomorphism)  to  the $\text{SO}(3)$ group. For spin~$1$, the matrices are unitarily similar to $\text{SO}(3)$, and they can be  transformed into  generators of  $\text{SO}(3)$ via  unitary  transformations.

Two observations, which we describe next,  lead us to adopting the Geometric Quantization (GQ) method. 
One is the  relevance of the $\text{SO}(3)$ group to modeling spin states which leads to the quantizing of the sphere itself. The other is that at any given time, a spin observable is the spin value along one chosen direction, say $z$ direction,  and  thus the uncertainty is between the $z$ direction and {\rm not-the-}$z$ direction.   

\subsection{The Geometric Quantization Approach to  Spin Phase Space}
\label{sec:Geometric-Quantization}
The phase space of a spin is derived from quantizing the sphere as it is developed  by the GQ method, e.g.,  see  \cite{woodhouse1997geometric,blau1992symplectic, nair2016elements}, and we summarize it now. The sphere is the surface of the ball with a radius of the spin magnitude $s \hbar$.

On the sphere, the $z$ values are specified by the polar representation  $s \hbar \cos \theta$, while the values on the intersection with  planes perpendicular to the $z$ axis are specified by the azimuth angle $\phi$. Treating the sphere as a phase space, one assigns  a rescaled  simpletic 2-form in spherical polar coordinates $ \diff \omega = \diff p \wedge \diff q= s \hbar \sin \theta \diff \theta \wedge \diff \phi$, and so  the Lagrangian is ${\cal L}=p \dot q  =s \hbar  \cos \theta \, \dot \phi$, and  the action is $S_{\cal  L}=\int\! {\cal L} \diff t = s \hbar  \int \!\cos \theta  \diff \phi $.
From this Lagragian, the GQ method derives the uncertainty commutation relation
\begin{align}
[\phi, s \hbar \cos \theta ] &= \iu \hbar \, ,
\label{eq:spin-commutation-GQ}
\end{align}
where the conjugate pair of eigenvalues $(\phi, s \hbar \cos \theta)\in [0,2\piu) \times   \{-s\hbar,\hdots, s\hbar\}$ forms a spin phase space with a finite Hilbert space volume $2 (2s+1)\piu  \hbar$. 

Note that the rotation operator $\eu^{-\iu  \frac{S_z}{\hbar}\phi}$ of an angle $\phi$ around the $z$ axis describes the polarization angle in the $x$-$y$ plane. Thus we will refer to the angle $\phi$ as the polarization angle. In order to create physical quantities with the polarization operator $\phi$, we must constrain $\phi$ to a periodic function  with period $2\pi$, i.e., to values that are a function of $\eu^{\iu \phi}$.

At the northern pole ($\cos \theta=1$) and the southern pole ($\cos \theta=-1$) the angles $\phi$ are not  defined. Thus,  in the basis $\ket{\phi}$ that diagonalizes $\phi$   we have  two  operators, namely  $  s \hbar \cos \theta =s\hbar-\iu\hbar\frac{\partial}{\partial \phi}$ for  the northern hemisphere and   $s \hbar \cos \theta =-s \hbar -\iu\hbar\frac{\partial}{\partial \phi}$  for the southern hemisphere.

In the basis $\ket{\phi}$, the eigenstates of the operator $S_z=s \hbar \cos \theta$ are
\begin{align}
    \ket{\xi_{s,m}}= \int  \ket{\phi} \bra{\phi}\ket{\xi_{s,m}}\diff \phi=\int \psi_{s,m}(\phi) \ket{\phi} \diff \phi \,,
\end{align}
where $m=-s,\hdots, s$, and
\begin{align}
    \psi_{s,m}(\phi)=\left \{ \begin{matrix}
    \frac{1}{\sqrt{2\piu}} \eu^{\iu  (s+m) \phi}, & \, m\ge 0 &\qquad \text{(northern hemisphere)}\, ;
    \\
    \frac{1}{\sqrt{2\piu}} \eu^{\iu  (-s+m) \phi},& \, m <0  &\qquad \text{(southern hemisphere)}\, .
    \end{matrix}\right.
    \label{eq:psism}
\end{align}
The two solutions in \eqref{eq:psism} are periodic in $\phi$ and differ by a phase (gauge) transformation of $\eu^{-\iu  2 s \phi}$. 

Consider a particle state $\ket{\xi_{s}}$ of spin magnitude $s \hbar$.  This state in the basis of the eigenvectors of  $S_z=s \hbar \cos \theta$ and $S^2$ is
\begin{align}
      \ket{\xi_{s}}=\sum_{m=-s}^s \alpha_{s,m} \ket{\xi_{s,m}}\,,
\end{align}
with $1=\sum_{m=-s}^{s}|\alpha_{s,m}|^2$. In the basis of the conjugate variable $\phi$,  the state is
\begin{align}
     \ket{\xi_{s}} & =  \int_0^{2\piu} \ket{\phi}\bra{\phi}\ket{\xi_{s}} \diff \phi  = \int_0^{2\piu} \ket{\phi} \sum_{m=-s}^s \alpha_{s,m} \bra{\phi}\ket{\xi_{s,m}}  \diff \phi
     \\
     &= \int _0^{2\piu} \sum_{m=-s}^s \alpha_{s,m} \psi_{s,m}(\phi) \ket{\phi} \diff \phi
     = \int_0^{2\piu} \lambda_{s}(\phi) \ket{\phi} \diff \phi
     \label{eq:xi-phi}\,,
\end{align}
where 
\begin{align}
    \lambda_{s}(\phi)= \bra{\phi}\ket{\xi_{s}}=\sum_{m-s}^s \alpha_{s,m} \psi_{s,m}(\phi)
    \label{eq:lamda-phi}\,.
\end{align}
Thus, for a  state $\ket{\xi_{s}} $ with density matrix $\rho_{s}= \ket{\xi_{s}} \bra{\xi_{s}} $   the  probabilities of the phase space are the product of the probabilities $ \{\rho_{s,m}=\bra{\xi_{s,m}}\rho_{s}\ket{\xi_{s,m}}=|\alpha_{s,m}|^2\}$ with the probability densities $\{ \rho_s(\phi)=\bra{\phi}\rho_{s}\ket{\phi}=|\lambda_{s}(\phi) |^2\}$.

\subsection{The Spin-Entropy in Spin Phase Space}
 
Our goal is to quantify the randomness given  a spin state. In order to capture  the randomness of the observables without double counting them, we specify the spin conjugate operators that creates the phase space. The phase space  captures all the randomness of the observables of a spin state. 
 
We define the  spin-entropy of a pure quantum state $\ket{\xi_{s}} $ with spin $s$ in spin phase space to be
\begin{align}
    \entropyS=\entropyS^{z}+\entropyS^{z^{\perp}}&=-\sum_{m=-s}^s\rho_{s,m} \ln \rho_{s,m}-  \int \!  \rho_{s}(\phi) \ln   \rho_{s}(\phi) \, \diff \phi\,.\\
    &=-\sum_{m=-s}^s|\alpha_{s,m}|^2\ln |\alpha_{s,m}|^2-  \int \! |\lambda_{s}(\phi)|^2 \ln |\lambda_{s}(\phi)|^2 \, \diff \phi\,.
    \label{eq:spin-entropy-geometric-quantization}
\end{align}
The first  term is the Shannon entropy  capturing the randomness of the spin value along the  $z$ axis. The second   term is  differential entropy  capturing the randomness of the spin value in the plane perpendicular to the $z$ axis, i.e., the entropy of the polarization angle $\phi$.   We define \emph{intrinsic-spin-information}, represented by   $\upGamma$, as the inverse of the entropy, i.e.,
\begin{align}
    \upGamma =\frac{-1}{\sum_{m=-s}^s|\alpha_{s,m}|^2\ln |\alpha_{s,m}|^2+  \int \! |\lambda_{s}(\phi)|^2 \ln |\lambda_{s}(\phi)|^2 \, \diff \phi}\,.
    \label{eq:knowledge-geometric-quantization}
\end{align}
$\upGamma$ is  non-negative and quantifies how much it is known about the observables of a state, where infinity indicates full knowledge of all observable values simultaneously.  

We conjecture that the spin-entropy \eqref{eq:spin-entropy-geometric-quantization} depends only on the variables that define the spin-entropy's component along the $z$ axis. This is further elaborated  and utilized in Conjecture~\ref{conj:z-dependence}. We also conjecture that the ordering (ranking) of the states according to the spin-entropy can be obtained solely from  the spin-entropy component along the $z$ axis. This conjecture implies that states with maximum and minimum spin-entropy values are the ones with maximum and minimum spin-entropy component along the $z$ axis.

\subsection{Illustrative Example}
To illustrate, consider the case of a state with $\alpha_{s,m}=\sqrt{\nicefrac{2\pi}{2s+1}}\, \psi^*_{s,m}(\phi_0)$. Then, we get the  uniform distribution $|\alpha_{s,m}|^2=\nicefrac{1}{2s+1}$, i.e., thus a  state with the lowest intrinsic-spin-information  along the  $z$ axis.   Then
\begin{align}
    \lambda_{s}(\phi) &= \frac{1}{\sqrt{2\pi\, 2s+1}}   \sum_{m=-s }^s \eu^{\iu  \left((2\theta(m)-1)\, s+m \right) (\phi-\phi_0)}
    \\
    &= \frac{1}{\sqrt{2\pi\, (2s+1)}}\Big ( 1-\Mod(2s,2)+ 2 \smashoperator{\sum_{j=j_{\min}}^{2s}} \cos \big(j\,(\phi-\phi_0)\big) \Big)\, ,
\end{align}
where $j_{\min}=s+1-\frac{1}{2}\Mod(2s,2)$, and $\Mod$ is the modulus function. Note that the distribution $\lambda_{s}(\phi) $ becomes more concentrated around $\phi_0$  as $s$ increases.  The intrinsic-spin-information about the spin  on the plane $z^{\perp}$ increases as $s$ increases. However, the total intrinsic-spin-information reduces because of the increase in spin-entropy along $z$. For a fermion of spin \nicefrac{1}{2} the spin-entropy  becomes 
\begin{align}
    \entropyS&=\ln2+\ln \pi-  \frac{1}{\pi}\int_0^{2\pi} \! \cos^2 (\phi-\phi_0) \ln \cos^2 (\phi-\phi_0) \, \diff \phi\\
    &=\ln 2 +\ln \pi+\ln \frac{4}{e}\approx 1.531+\ln 2\, ,
    \label{eq:spin-entropy-geometric-quantization-electron}
\end{align}
and $\upGamma\approx 0.450$. For spin $1$, $\entropyS\approx1.270+\ln 3$ with $\upGamma\approx 0.422$.

\subsection{Minimum Spin-Entropy }

The third law of thermodynamics establishes zero as the minimum of thermodynamics entropy. The use of differential entropy for the term $\entropyS^{z^{\perp}}_{s,\phi}$ may suggest  that  the proposed entropy could be negative, but  we will establish a positive minimum  for it.

In quantum mechanics, ignoring the spin and focusing on the spatial DOFs, the entropic uncertainty principle~\cite{hirschman1957note,beckner1975inequalities,  bialynicki1975uncertainty},  establishes that $\entropyS_x+\entropyS_p \ge 3  \ln(\eu \, \hbar\, \piu)$, with equality  for the normal distribution.  A similar bound  can now be derived for the spin-entropy.

\begin{lemma}[Hausdorff-Young Inequality Holds]
\label{lemma:Hausdorff-Young-inequality}
The Hausdorff-Young inequality holds for the spin states phase space, i.e., for  $p\in (1,2]$ and for every $q$ satisfying $1=\frac{1}{p}+\frac{1}{q}$,
\begin{align}
  \Big ( \sum _{m=-s }^{s }\big |\alpha_{s,m}\big |^{q}\Big )^{1/q}  &\le \bigg ((2\piu)^\frac{p-2}{2}\int_0^{2\piu}|\lambda_{s}(\phi)|^{p}\,\diff \phi\bigg )^{1/p}\, .
\label{eq:Hausdorff-Young-inequality}
\end{align}
\end{lemma}
\begin{proof}
Rewrite \eqref{eq:lamda-phi} as
\begin{align}
    \lambda_{s}(\phi)= \sum_{m=-\infty}^{\infty} \alpha_{s,m} \psi_{s,m}(\phi)\, .
    \label{eq:lamda-phi-II}
\end{align}
where $\alpha_{s,m}=0$ for $|m|> s$.
Note that the sum on $m$ is either on all  even  or all odd  multiples of $\frac{1}{2}$, depending on whether $2s$ is  even or odd. Then, we rewrite  \eqref{eq:lamda-phi-II} as
\begin{align}
    \lambda_{s}(\phi)& =\frac{1}{\sqrt{2\piu}}\sum_{m=-\infty}^{\infty} \alpha_{s,m}\, \eu^{ \iu [(2\theta(m)-1)\, s+m] \phi}
    =\frac{1}{\sqrt{2\piu}}\sum_{j=-\infty}^{\infty} \beta_{s,j} \, \eu^{ \iu j \phi} \,,
    \label{eq:lamda-phi-Fourier}
\end{align}
where $j \in \mathbb{Z}$, $\beta_{s,j}=\alpha_{s,m}$ for $j=\big (2\theta(m)-1\big )\, s+m$  and $\beta_{s,j}=0$ otherwise.    Clearly, for any value of $r$ 
\begin{equation}
  \sum_{j=-\infty}^{\infty} |\beta_{s,j}|^r=\sum_{m=-\infty}^{\infty} |\alpha_{s,m}|^r  \, .
  \label{eq:beta-alpha-sum}
\end{equation} 
Then \eqref{eq:lamda-phi-Fourier} describes  a Fourier series expansion of the periodic complex-valued function $ \lambda_{s}(\phi)$. The Fourier inverse is
\begin{align}
   \frac{1}{\sqrt{2\piu}}\int_0^{2\piu} \lambda_{s}(\phi) \, \eu^{- \iu j \phi} \diff \phi =\beta_{s,j}\qquad
  {\rm or}\qquad  \int_0^{1} \sqrt{2\piu} \, \lambda_{s}(2\piu \mu) \, \eu^{- 2\piu \iu j  \mu} \diff \mu =\beta_{s,j}\, .
    \label{eq:lamda-phi-Fourier-inverse}
\end{align}
Thus,  the Hausdorff-Young inequality
\begin{align}
    \Big ( \sum_{j=-\infty}^{\infty}\big |\beta_{s,j}\big |^{q}\Big )^{1/q}
    &\leq
     \bigg ( \int_0^{1} \left |\sqrt{2\piu} \, \lambda_{s}(2\piu \mu)\right|^p \,  \diff \mu \bigg )^{1/p}
     \label{eq:Hausdorff-Young-II}
\end{align}
     holds. We can then replace back   the variables $\alpha_{s,m}$'s and $\phi=2\piu \mu$ into \eqref{eq:Hausdorff-Young-II}, completing the proof.
\end{proof}

\begin{theorem}
\label{th:minimun-spin-entropy}
The spin-entropy  satisfies the inequality
\begin{align}
    \entropyS\ge \ln 2\piu\, ,
    \label{eq:min-spin-entropy}
\end{align}
with equality  attained for the eigenstates of the spin operator $S_z=s \hbar \cos \theta$.
\end{theorem}
\begin{proof}
This proof is an adaptation to the specific spin phase space of  the work of \cite{hirschman1957note,beckner1975inequalities,  bialynicki1975uncertainty}. 

First, by  Lemma~\ref{lemma:Hausdorff-Young-inequality}  we can apply  Beckner's theorem \cite{beckner1975inequalities}
\begin{align}
   \left (\frac{1}{q}   \right)^{\frac{n}{2q}}\, \left (\frac{1}{p}   \right)^{-\frac{n}{2p}} \ge  \frac{\Big ( \sum _{m=-s }^{s }\big |\alpha_{s,m}\big |^{q}\Big )^{1/q}}{\bigg ( (2\piu)^\frac{p-2}{2}\int_0^{2\piu}|\lambda_{s}(\phi)|^{p}\,\diff \phi\bigg )^{1/p}}\,,
\label{eq:Beckner}
\end{align}
where $1 < p\le 2$, $\frac{1}{p}+\frac{1}{q}=1$,   and $n$ is the dimensionality of the space in which the functions are defined. For the sphere,  $n=2$. Applying  the logarithm to both sides of the inequality and noting that all the quantities are non-negative, we get
\begin{align}
  \ln \left[ \left (\frac{1}{q}   \right)^{\frac{1}{q}}\, \left (\frac{1}{p}   \right)^{-\frac{1}{p}} \right]+\ln \bigg ((2\piu)^\frac{p-2}{2}\int_0^{2\piu}|\lambda_{s}(\phi)|^{p}\,\diff \phi\bigg )^{1/p} \ge  \ln \Big (\sum _{m=-s }^{s }\big |\alpha_{s,m}\big |^{q}\Big )^{1/q}\,.\quad
\label{eq:log-beckner}
\end{align}
Substituting  $p=2\alpha$ and $q=2\beta$, we get the constraint $ \nicefrac{\alpha}{1-\alpha}=-\nicefrac{\beta}{1-\beta} \in (1,\infty)$ from $1=\nicefrac{1}{p}+\nicefrac{1}{q}$. Multiplying the  terms on both sides of  \eqref{eq:log-beckner} by  either one of these equal and positive expressions $ \nicefrac{\alpha}{1-\alpha}$ or $-\nicefrac{\beta}{1-\beta}$, and  rearranging terms, we  get
\begin{align}
 \frac{1}{1-\alpha} \ln \bigg (\int_0^{2\piu}|\lambda_{s}(\phi)|^{2\alpha}\,\diff \phi\bigg ) +\frac{1}{1-\beta} \ln \Big (\sum _{m=-s }^{s }\big |\alpha_{s,m}\big |^{2\beta}\Big )
 \\
 \quad \ge \frac{1}{1-\alpha} \ln  \left (\frac{1}{2\alpha}   \right) +\frac{1}{1-\beta}\ln \left (\frac{1}{2\beta}   \right)+\ln 2\pi \,,
\label{eq:log-beckner-Renyi}
\end{align}
where the two first terms are the  R\'enyi's differential entropy and the R\'enyi's  entropy, forming an entropic inequality. Thus, by taking the limits $\alpha \rightarrow 1^{-}$ and $\beta \rightarrow 1^{+}$, we  obtain  \eqref{eq:min-spin-entropy}.

The minimum entropy  is reached for the eigenstates of $S_z$, $\ket{\xi_{s,z}}=\ket{\xi_{s,m_0}}$, for any eigenvalue $\hbar m_0$ where $-s\le m_0 \le s$. Then $\alpha_{s,m}=\updelta_{m,m_0}$, and so $\lambda_{s}(\phi)=\psi_{s,m_0}$. The probabilities are then  $|\alpha_{s,m}|^2=\updelta_{m,m_0}$, and so the Shannon entropy is zero. The  probability density is then $|\lambda_{s}(\phi)|^2= \nicefrac{1}{2\piu} $, with differential  entropy $\ln 2\piu$.
\end{proof}
Thus, for the intrinsic-spin-information  $0\le \upGamma \le \nicefrac{1}{\ln 2\pi}$.

Two observations: 

\begin{enumerate}
\item The spin-entropic principle inequality for  a bounded variable $\phi$ and an integer variable $m$ resulting from   the Fourier series transformation reaches its minimum for  the  eigenstates of the $S_z, S^2$ operators, with corresponding $S_z$ eigenvalues  $m$'s.

\item Preparing a system  to align a spin state with a particular direction, an eigenstate $\ket{\xi_{s,m=s}}$ of a  $z$ axis,   as it is done in many experiments with spin up along $z$, provides the knowledge of  the spin  along that direction. Thus, the intrinsic-spin-information is gained by such a  preparation of the lowest spin-entropy state.
\end{enumerate}

\subsection{Extending the Spin-entropy to Mixed States}

Mixed states extend the Hilbert space of specified quantum states, or pure states, to quantum states that are not fully specified and instead are described by a classical probabilistic combination of pure states.  Let a spin mixed state be defined via the density matrix 
\begin{align}
    \rho^{\rm M}_s=\sum_{i=1}^N\gamma_i \ket{\xi_{s}^i}\bra{\xi_{s}^i}\, ,
\end{align}
where $\ket{\xi_{s}^i}=\sum_{m=-s}^s \alpha^i_{s,m} \ket{\xi_{s,m}},\,  i=1,\hdots, N$ are pure states. 
The randomness present in a pure state is concentrated in the  observables of the spin phase space.  However, for mixed states additional  randomness exists, that of specifying the state itself and is captured by the probabilities $\gamma_i$. Then, the distributions after a choice of  $z$ axis are
\begin{align}
\rho^{\gamma}_{s,m,i}&= \gamma_i\, \bra{\xi_{s,m}}\ket{\xi_{s}^i}\bra{\xi_{s}^i}\ket{\xi_{s,m}}=\gamma_i\,  |\alpha^i_{s,m}|^2 \, ,
\\
\rho^{\gamma}_{s,m}&= \bra{\xi_{s,m}}\rho^M\ket{\xi_{s,m}}=\sum_{i=1}^N\rho^{\gamma}_{i, s,m}=\sum_{i=1}^N\gamma_i\,  |\alpha^i_{s,m}|^2\, ,
    \label{eq:two-distributions-pure-mixed-states}
\end{align}
with the normalizations $1=\sum_{i=1}^N\sum_{m=-s}^s\rho^{\gamma}_{i, s,m}$ and $1=\sum_{m=-s}^s\rho^{\rm M}_{s,m}$. Thus, extending  \eqref{eq:spin-entropy-geometric-quantization} to include  mixed states, the spin-entropy is then
\begin{align}
    \entropyS^{\text{M}}&=-\sum_{i=1}^N \sum_{m=-s}^s\rho^{\gamma}_{s,m,i} \ln \rho^{\gamma}_{s,m,i}-  \sum_{i=1}^N \int \!   \rho^{\gamma}_{s,i}(\phi) \ln   \rho^{\gamma}_{s,i}(\phi)\, \diff \phi\,.
    \\
    & =S^{\text{vN}}+\sum_{i=1}^N \gamma_i \entropyS^i_s\, ,
    \label{eq:spin-entropy-geometric-quantization-mixed-states-hybrid}
\end{align}
where \vN entropy is given by $S^{\text{vN}}=-\sum_{i=1}^N\gamma_i \ln \gamma_i $ and  $\entropyS^i_s=-\sum_{m=-s}^s|\alpha^i_{s,m}|^2 \ln |\alpha^i_{s,m}|^2+\int   \lambda_{s}^i(\phi) \ln   \lambda_{s}^i(\phi)\, \diff \phi$ is the entropy of each pure state in the mixture. This entropy is always larger than \vN entropy, since we also consider the randomness of the observables and for every $i$, $ \entropyS^i_s\ge 0$, and so $\sum_{i=1}^N \gamma_i \entropyS^i_s \ge 0$.

Note that if one focuses only on the randomness of the observables, the distributions in spin phase space are 
\begin{align}
   \rho^{\rm M}_{s,m}&= \bra{\xi_{s,m}}\rho^M\ket{\xi_{s,m}}=\sum_{i=1}^N\gamma_i\,  |\alpha^i_{s,m}|^2 \, ,
   \\
  \rho^{\rm M}_s(\phi) &= \bra{\phi}\rho^M\ket{\phi}=\sum_{i=1}^N\gamma_i \left( \sum_{m'=-s}^s \alpha^i_{s,m'}\psi_{s,m'}(\phi)\right)^* \sum_{m=-s}^s \alpha^i_{s,m}\psi_{s,m}(\phi)\, .
\end{align}

\subsection{Pure System with Multiple Particles  and Quantum Field Theory}

Consider the spin of a pure system with $N$  particles of the same species, each with spin $s$. The maximum spin  $s_{\max}$ of the system occurs when all  the $N$ particles are aligned  yielding  $s_{\max}= Ns$.  For a state with spin value $s_{\max}$, the possible $z$ values are   $m=-s_{\max}, \hdots, s_{\max}$. However, for the  system with $N$  particles all different values $s'\in [s_{\min}, s_{\max}]$ can occur, where $s_{\min}=s\,\Mod(N,2)$ for fermions and $s_{\min}=0$ for bosons. For each  $s'$,  all possible $z$ components must be considered, so $m_{s'}=-s',\hdots,s'$.
The eigenstate basis of the  $S_z, S^2$ operators associated with the system is then
\begin{align}
   \left \{\ket{\xi^{N,s}}\right \}&=\cup_{s'=s_{\min}}^{s_{\max}} \cup_{m=-s'}^{s'} \ket{\xi_{s',m}}\, .
\end{align}
For example, two fermions with $s=\nicefrac{1}{2}$ will produce four $z$ states, namely three states with  $s'=1$ and $m'=-1,0,1$, and the fourth state with $s'=0  $ and $ m'=0$. Thus, the eigenstate basis of $S_z, S^2$ operators for two fermions   is $\left \{\ket{\xi_{1,1}},\ket{\xi_{1,0}},\ket{\xi_{1,-1}},\ket{\xi_{0,0}}\right \}$.  
In order to extend the spin-entropy to $N$ particles of the same species, we first describe  the density matrix of a state to be
\begin{align}
\rho^{N,s}&=\ket{\xi^{N,s}}\bra{\xi^{N,s}}=\left(\sum_{s'=s_{\min}}^{s_{\max}}\sum_{m_{s}=-s'}^{s'}\alpha_{s',m_{s}}\ket{\xi_{s',m_{s}}}\right)\left(\sum_{s''=s_{\min}}^{s_{\max}}\sum_{m'_{s}=-s''}^{s}\alpha_{s'',m'_{s}}\bra{\xi_{s'',m'_{s}}}\right)^*
\\
&=\sum_{s'=s_{\min}}^{s_{\max}}\sum_{m_{s}=-s'}^{s'}\sum_{s''=s_{\min}}^{s_{\max}}\sum_{m'_{s}=-s''}^{s''}\alpha_{s',m_{s}}\alpha^*_{s'',m'_s} \ket{\xi_{s',m_{s}}}\bra{\xi_{s'',m'_{s}}}\, .
    \label{eq:density-matrix-N-spin-s}
\end{align}
Thus, projecting \eqref{eq:density-matrix-N-spin-s} onto the $z$-basis and onto the $\ket{\phi}$ basis we derive the density functions in spin phase space  for $s'\in [s_{\min},s_{\max}]$ as
\begin{align}
\rho^{N,s}_{m}=\bra{\xi_{s',m}}\rho^{N,s}\ket{\xi_{s',m}}&=|\alpha_{s',m}|^2\,, \qquad \text{for}  \quad m\in [-s',s'], \quad \text{and}
\\
\rho^{N,s}(\phi)=\bra{\phi}\rho^{N,s}\ket{\phi}&= |\lambda_{s'}(\phi)|^2 =\sum_{m=-s'}^{s'} \sum_{m'=-s'}^{s'}\alpha_{s',m}\alpha^*_{s',m'} \, \psi_{s',m}(\phi)  \psi^*_{s',m'}(\phi)\,.
    \label{eq:density-matrix-N-spin-s-phase-space}
\end{align}
Then the spin-entropy \eqref{eq:spin-entropy-geometric-quantization} of the system is
\begin{align}
    \entropyS_{N,s}&=\entropyS^{z}_{N,s}+\entropyS^{z^{\perp}}_{N,s}
    \\
    &=-\sum_{s'=s_{\min}}^{s_{\max}}\sum_{m=-s'}^{s'}\rho^{N,s}_{m}\ln \rho^{N,s}_{m}-  \sum_{s'=s_{\min}}^{s_{\max}}\int \rho^{N,s}(\phi)\ln \rho^{N,s}(\phi) \, \diff \phi
    \\
    &=-\sum_{s'=s_{\min}}^{s_{\max}}\left (\sum_{m=-s'}^{s'}|\alpha_{s',m}|^2\ln |\alpha_{s',m}|^2-  \int |\lambda_{s}(\phi)|^2 \ln |\lambda_{s}(\phi)|^2 \, \diff \phi\right)\, .
    \label{eq:spin-entropy-geometric-quantization-multiple-particles}
\end{align}
where the normalization is  $1=\sum_{s'=s_{\min}}^{s_{\max}}\sum_{m=-s'}^{s'}\rho^{N,s}_{m}$ and $1=\sum_{s'=s_{\min}}^{s_{\max}}\int |\lambda_{s}(\phi)|^2 \, \diff \phi$, and the phase space to describe a system of $N$ particles of spin $s$ consists of all  the spheres with radiuses $\hbar s$, for $s \in [s_{\min}, s_{\min}+1, \hdots, s_{\max}-1, s_{\max}]$.

In quantum field theory (QFT), superpositions of  states with any number of particles of spin species s are also considered. One can write a state as
\begin{align}
\ket{\xi_{\QFT}}=\sum_{N=1}^{\infty}\gamma_N \sum_{s'=s_{\min}}^{s_{\max}}\sum_{m=-s'}^{s'}\alpha^N_{s',m} \ket{\xi_{s',m}}\, ,
\end{align}
where $\sum_{N=1}^{\infty}|\gamma_N|^2=1$ and both, $s_{\min} $ and $ s_{\max}$, depend on $N$ and $s$. Due to the orthogonality of the states with different number of particles,   it is straightforward to extend \eqref{eq:spin-entropy-geometric-quantization-multiple-particles} to obtain
\begin{align}
    \entropyS_{\QFT,s}&=\sum_{N=1}^{\infty}|\gamma_N|^2\left(\entropyS^{z}_{N,s}+\entropyS^{z^{\perp}}_{N,s}\right) -\sum_{N=1}^{\infty}|\gamma_N|^2\ln |\gamma_N|^2 
    \, .
    \label{eq:spin-entropy-geometric-quantization-QFT}
\end{align}
Thus, in QFT the role of the magnitude square of the complex valued coefficient $\gamma_N$ resembles the mixed  states coefficients to the entropy \eqref{eq:spin-entropy-geometric-quantization-mixed-states-hybrid} .

\subsection{The {\em z} axis}
The formulation of a spin operator and corresponding eigenstates requires a choice of a $z$ axis. This is  evident from the spin phase space where the quantization of the spin is along the $z$ axis, and perpendicular to it a continuous polarization variable $\phi\in[0,2\pi)$ is assigned.   A choice of $z$ axis must be made to construct the quantum state and assign to it a unique entropy. Approaches such as \cite{lieb2014proof} average  their proposed entropy over all possible 3D rotations to eliminate the $z$ axis bias, ending up with a spin-entropy that bundles a large set of quantum states into one value. Instead, we investigate a physical property that causes the break of the isotropy of the 3D space for spin states. For  the Stern–Gerlach (SG) experiment, the direction of the applied magnetic field does define the $z$ axis, splitting spins according to a positive $z^+$ or negative $z^-$ direction eigenstates.

For a system of particles with total spin $S$ its spin Hamiltonian can be described by $H= \frac{q \hbar}{2m} (g B_{\rm e} + (g-1) B_{\rm I})\cdot  S $ where $``\cdot "$ is the scalar product, $g$ is the gyromagnetic ratio, $q$ the charge of the system, $m$ its inertial mass, $B_{\rm e}$ is the external magnetic field applied to the spin system, and $B_{\rm I}=\frac{1}{c}\nicefrac{E\times v}{\sqrt{1-\frac{v^2}{c^2}}}$ is the magnetic field  in the rest frame of the system  when the charged system is moving in an electric field $E$ with velocity $v$.  Thus, the preferred direction that breaks the isotropy of the 3D space and defines the $z$ axis is  $g B_{\rm e} + (g-1) B_{\rm I}$. 
 
Consider the example of  photon emission by an excited hydrogen atom in state  $2p_z (n=2, l=1, m=1)$ transitioning to state $1s (n=1, l=0, m=0)$. The  $z$ axis defines the quantum state associated with the numbers $m=1,0,-1$, i.e., the excited state already broke the isotropy  and consequently  the angular momentum conservation leads to the emission of a photon on a plane perpendicular to this $z$ axis.

\section{Spin-Entropy for One Particle}

We first analyze  the spin-entropy for a fermion with spin value $\nicefrac{1}{2}$,  then we analyze a massive boson with spin $1$, and we conclude this section by analyzing the photon entropy.

\subsection{Spin \nicefrac{1}{2}}
\begin{figure}
 \centering
 \includegraphics[scale=0.3]{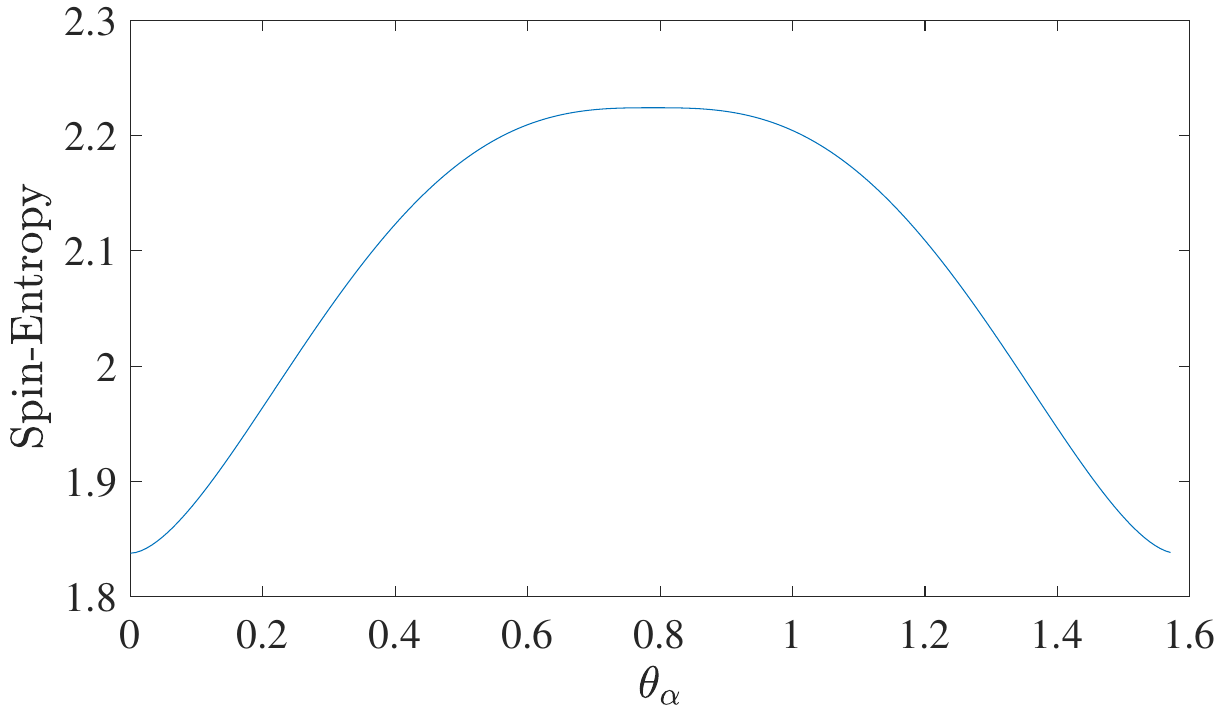}
 \caption{A plot of spin-entropy \eqref{eq:entropy-spin-half}  vs $ \theta_{\alpha} \in [0,\frac{\piu}{2}]$, for $s=\nicefrac{1}{2}$.}
 \label{fig:spin-entropy}
\end{figure}

A spin state of a particle with spin $\nicefrac{1}{2}$,  represented by  a set of two orthonormal eigenstates
$\ket{+} =\ket{\xi_{\nicefrac{1}{2},\nicefrac{1}{2}}}= (1,0)^{\tran}$ and $\ket{-} =\ket{\xi_{\nicefrac{1}{2},-\nicefrac{1}{2}}}= (0,1)^{\tran}$
of the operators $(S^2,S_z)$ with associated eigenvalues $\big(s=\nicefrac{1}{2},m=\pm \nicefrac{1}{2}\big)$, is
\begin{align}
    \ket{\xi_{\nicefrac{1}{2},z}}=\eu^{\iu \varphi}\left ( \eu^{\iu \nu}\cos \theta_{ \alpha} \ket{+}+\sin \theta_{ \alpha} \ket{-}\right)=\eu^{\iu \varphi}\begin{pmatrix}
\eu^{\iu \nu}\cos \theta_{ \alpha}, & \sin \theta_{ \alpha}
\end{pmatrix}^{\tran}
\label{eq:spinHalfstate}
\end{align}
with  $\theta_{\alpha} \in [0,\frac{\piu}{2}]$ and  $\varphi, \nu \in [0,\piu)$.
\begin{proposition}[spin-entropy for $s=\nicefrac{1}{2}$]
\label{prop:entropy-spin-half}
The spin-entropy of a spin state with $s=\nicefrac{1}{2}$,  described by \eqref{eq:spinHalfstate},  is
\begin{align}
    \entropyS_{\frac{1}{2}}(\theta_{ \alpha} )&=- \cos^2 \theta_{\alpha} \ln \cos^2 \theta_{ \alpha} - \sin^2 \theta_{\alpha} \ln \sin^2 \theta_{ \alpha}+\ln 2\piu
    \\
    &\quad -\frac{1}{2\piu} \int_0^{2\piu} \left ( 1+  \sin 2 \theta_{ \alpha} \cos 2 \phi\right) \ln \left ( 1+  \sin 2 \theta_{ \alpha} \cos 2 \phi\right)\diff \phi \, ,
    \label{eq:entropy-spin-half}
\end{align}
\end{proposition}
\begin{proof}
A state  $\ket{\xi_{\nicefrac{1}{2}}}$ assigns the probability distribution along the $z$ axis eigenstates
$\pr_{z}  =\begin{pmatrix}
\left|\bra{+}\ket{\xi_{\nicefrac{1}{2}}}\right|^2, &
\left|\bra{-}\ket{\xi_{\nicefrac{1}{2}}}\right|^2
\end{pmatrix}^{\tran}=  \begin{pmatrix}
\cos^2 \theta_{ \alpha}, &
\sin^2 \theta_{ \alpha}
\end{pmatrix}^{\tran}$.
Thus, from \eqref{eq:xi-phi} and \eqref{eq:lamda-phi} we get $\ket{\xi_{\nicefrac{1}{2},\phi}}=\int \eu^{\iu \varphi}\frac{1}{\sqrt{2\piu}}\left(
\eu^{\iu \nu}\eu^{\iu \phi} \cos \theta_{ \alpha} +   \eu^{-\iu \phi} \sin \theta_{ \alpha}\right) \ket{\phi}  \diff \phi $, and so $\rho_{\theta_{\alpha},\nu}(\phi)=\frac{1}{2\piu}(1+  \sin 2 \theta_{ \alpha} \cos (2 \phi+\nu))$.
\end{proof}
For a visualization of the entropy, see Figure~\ref{fig:spin-entropy}.

Note the Wehrl's entropy for any state $\ket{\xi_{\nicefrac{1}{2}}}$ is a constant since all spin \nicefrac{1}{2} states  are the same coherent state up to a phase \cite{lieb:91, lieb2014proof},  yielding the same entropy. 

\subsection{Spin 1}
For massive particles with spin $s=1$,  the spin matrices are
\begin{align}
 S_x =\frac{1}{\sqrt{2}} \begin{pmatrix}      0 & 1 & 0    \\    1 & 0  &  1    \\        0 & 1  & 0   \end{pmatrix} \,,\quad     S_y =\frac{1}{\sqrt{2}}\begin{pmatrix}      0 & -\iu &  0    \\    \iu & 0 & - \iu    \\     0 & \iu &  0    \end{pmatrix}\,,
 \quad
 S_z =\begin{pmatrix}      1 & 0  & 0    \\   0 & 0  &  0    \\        0 & 0  & -1 \end{pmatrix}\,,   \quad   S^2 =\begin{pmatrix}     2 & 0  & 0    \\   0 & 2 &  0    \\        0 & 0 & 2 \end{pmatrix}\,,
\end{align}
yielding a basis representation formed with the eigenvectors of  $S_z$ and $S^2$:
$\ket{\uparrow}_z=\ket{\xi_{1,1}}=\begin{pmatrix}
    1,&  0,& 0
    \end{pmatrix}^{\tran}$, $\ket{\rightarrow}_z=\ket{\xi_{1,0}}=\begin{pmatrix}
    0, &  1, & 0
    \end{pmatrix}^{\tran} $, $ \ket{\downarrow}_z=\ket{\xi_{1,-1}}=\begin{pmatrix}
     0, & 0, & 1
    \end{pmatrix}^{\tran}$. A general state of spin $s=1$ in the basis aligned with the $z$ axis is
\begin{align}
\label{eq:state-spin-1-z}
  \ket{\xi_z} &=  \eu^{\iu \varphi_{y}}  \big(\sin\theta_{\alpha} \cos \theta_{\beta}\,  \eu^{\iu \varphi_{x}} \ket{\uparrow}_z  + \cos \theta_{\alpha}  \ket{\rightarrow}_z
              +\sin \theta_{\alpha} \sin \theta_{\beta}\, \eu^{\iu \varphi_{z}}    \ket{\downarrow}_z \big)\, ,
\end{align}
with  $\theta_{\alpha} , \theta_{\beta}  \in [0,\frac{\piu}{2}]$, and $\varphi_{x}, \varphi_{z}, \varphi_{y} \in [0,2\piu)$. Following \eqref{eq:xi-phi}, we write the state in the $\ket{\phi}$ basis as
 \begin{align}
\label{eq:state-spin-1-phi}
        \ket{\xi_{z^{\perp}}} & = \frac{\eu^{\iu \varphi_{y}}}{\sqrt{2\piu}}\int \big(\sin \theta_{\alpha} \cos \theta_{\beta}\,  \eu^{\iu (\varphi_{x}+2\phi)}  + \cos \theta_{\alpha}\eu^{\iu \phi}
              +\sin\theta_{\alpha} \sin \theta_{\beta}\, \eu^{\iu (\varphi_{z}-2\phi)} \big) \ket{\phi} \diff \phi.
\end{align}

\begin{figure}
 \centering
 \includegraphics[scale=0.4]{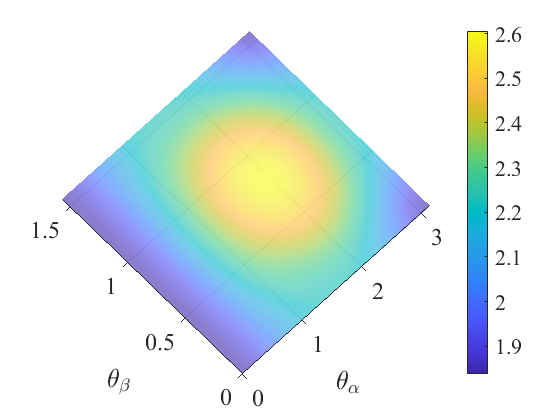}
 \caption{spin-entropy for $s=1$ \eqref{eq:spin1-entropy-conjecture} vs $(\theta_{\alpha}, \theta_{\beta})$. Note that for $\theta_{\alpha}=0$  and for  $\theta_{\alpha}=\frac{\piu}{2} $, $  \theta_{\beta}= 0,  \frac{\piu}{2}$, describing the three eigenstates along the $z$ axis,  the spin-entropy reaches its minimum value $1.838$, approximating $\ln 2\piu$.
}
 \label{fig:spin-entropy-GQ-1}
\end{figure}

\begin{proposition}
The  spin-entropy of a particle with spin  $s=1$ in a  state  given by \eqref{eq:state-spin-1-z}  is
\begin{align}
\label{eq:spin-1-entropy}
    \entropyS_{1} &=- \int_0^{2\piu} \rho_1(\phi,\varphi_{x},\varphi_z, \theta_{\alpha},\theta_{\beta} )\ln \rho_1(\phi,\varphi_{x},\varphi_z, \theta_{\alpha},\theta_{\beta} )\, \diff \phi\,
    \\
    &\quad +S_c(\cos^2\theta_{\alpha})+\sin^2\theta_{\alpha}\, S_c(\cos^2\theta_{\beta})\,  ,
\end{align}
where
\begin{align}
S_c(\cos^2\theta) &=-\cos^2 \theta\ln (\cos^2 \theta)-(1-\cos^2 \theta)\ln (1-\cos^2 \theta)
\\
\rho_1(\phi)&=\frac{1}{2\piu}\big|\sin \theta_{\alpha} \cos \theta_{\beta}\,  \eu^{\iu (\varphi_{x}+2\phi)}  + \cos \theta_{\alpha}\eu^{\iu \phi}
+\sin\theta_{\alpha} \sin \theta_{\beta}\, \eu^{\iu (\varphi_{z}-2\phi)} \big|^2
              \\
 &=\frac{1}{2\piu} \big[1+\sin^2 \theta_{\alpha} \sin (2\theta_{\beta}) \cos (4\phi+\varphi_x-\varphi_z)
             \\
&\quad  +\sin (2 \theta_{\alpha}) (\cos \theta_{\beta} \cos (\phi +\varphi_x)+\sin \theta_{\beta} \cos (3\phi-\varphi_z) \big ]\, .
\end{align}
\end{proposition}
\begin{proof}
Computing the three probabilities associated with state $\ket{\xi_z}$ and deriving its entropy term  $\entropyS^{z}_{1}$ yields
\begin{align}
    \entropyS^{z}_{1}
    &= -\cos^2 \theta_{\alpha}\ln \cos^2 \theta_{\alpha}-\sin^2\theta_{\alpha}\left [ \ln \sin^2\theta_{\alpha}-S_c(\cos^2\theta_{\beta})\right]\, .
\end{align}
The entropy term $\entropyS^{z^{\perp}}_{1}$ is derived  from the probability density associated with the state $\ket{\xi_{z^{\perp}}}$ from \eqref{eq:state-spin-1-phi}, and so
\begin{align}
    \rho_1(\phi)&=\frac{1}{2\piu}\big|\sin \theta_{\alpha} \cos \theta_{\beta}\,  \eu^{\iu (\varphi_{x}+2\phi)}  + \cos \theta_{\alpha}\eu^{\iu \phi}
              +\sin\theta_{\alpha} \sin \theta_{\beta}\, \eu^{\iu (\varphi_{z}-2\phi)} \big|^2\, .
\end{align}
\end{proof}
Observed in simulations and in  \eqref{eq:entropy-spin-half}, we  conjecture that
\begin{conjecture}
\label{conj:z-dependence}
The spin-entropy depends only on the variables that define the spin-entropy component along the $z$ axis.  In particular, the  spin-entropy  \eqref{eq:spin-1-entropy}   of a state  given by \eqref{eq:state-spin-1-z}  can be simplified to
\begin{align}
\label{eq:spin1-entropy-conjecture}
    \entropyS_{1}(\theta_{\alpha} , \theta_{\beta} )
    &=  - \int_0^{2\piu} \rho(\phi, \theta_{\alpha},\theta_{\beta} )\ln \rho(\phi, \theta_{\alpha},\theta_{\beta} )\, \diff \phi\,
    \\
    &\quad +S_c(\cos^2\theta_{\alpha})+\sin^2\theta_{\alpha}\, S_c(\cos^2\theta_{\beta}) \,  ,
\end{align}
where
\begin{align}
\rho(\phi, \theta_{\alpha},\theta_{\beta})&=\frac{1}{2\piu}\big|\sin \theta_{\alpha} \cos \theta_{\beta}\,  \eu^{\iu 2\phi}  + \cos \theta_{\alpha}\eu^{\iu \phi}
+\sin\theta_{\alpha} \sin \theta_{\beta}\, \eu^{-\iu 2\phi} \big|^2
\end{align}
\end{conjecture}
By Theorem~\ref{th:minimun-spin-entropy}, the  spin-entropy is minimized at $\theta_{\alpha}=0$, at $(\theta_{\alpha}=\nicefrac{\piu}{2}, \theta_{\beta}= 0)$, and at  $(\theta_{\alpha}=\nicefrac{\piu}{2}, \theta_{\beta}= \nicefrac{\piu}{2})$, the three eigenstates along the $z$ axis. Thus, preparing a spin  state orientation by aligning it with an axis, reduces the entropy. The entropy given by~\eqref{eq:spin1-entropy-conjecture} is visualized in Figure~\ref{fig:spin-entropy-GQ-1}.

In order to compare \eqref{eq:spin1-entropy-conjecture} to Wehrl's spin-entropy, we consider the spin $1$ coherent states
 \begin{align}
   \ket{s,\alpha}   &= \left (\cos \frac{\theta}{2}\right)^{2}  \ket{1,-1} +\sqrt{2} \cos \frac{\theta}{2} \sin \frac{\theta}{2} \eu^{\iu  \phi}\ket{1,0} + \left (\sin \frac{\theta}{2}\right)^{2} \eu^{\iu 2 \phi}\ket{1,1} \, .
   \label{eq:coherent-state-spin-1}
 \end{align}
The cases  where $\theta$ is $0$ or $\pi$ show that  the states  $\ket{1,-1}$ and $\ket{1,1}$ are coherent states, and thus minimize the Wehrl spin-entropy. However, state $\ket{1,0}$ cannot be a coherent state and therefore has higher Wehrl's spin-entropy. For our  spin-entropy, all eigenstates of $S_z$ minimimze the spin-entropy.  
Moreover, we note that coherent states \eqref{eq:coherent-state-spin-1} correspond to the general spin $1$ state \eqref{eq:state-spin-1-z} via the mapping 
\begin{align}
\phi=\varphi_{z}=-\varphi_{x} \, ,\quad     
\sin\theta=\sqrt{\frac{2\sin2 \theta_{\beta}}{1+\sin2 \theta_{\beta}}}=\sqrt{2}\cos \theta_{\alpha} 
\end{align}
There is a one-to-one map between $\theta$ and $2\theta_{\beta}$, both in the range $[0,\pi]$. Thus, changes in the angle $\theta_{\alpha}$ from the constraint  $\cos \theta_{\alpha}=\sqrt{\nicefrac{\sin2 \theta_{\beta}}{1+\sin2 \theta_{\beta}}}$ will increase Wehrl's entropy above its minimum.

\subsection{Photon Entropy}

Photon is a massless spin 1 particle. Its group representation is ISO(2) or E(2), with the gauge transformation accounting for the transverse direction.  Then, this results in a particle with helicity $\pm 1$ and a two dimensional polarization field in a plane propagating along, or against, the helicity direction. Formally one creates two circular polarized states, a right-hand side, $\ket{h_+}$, and a left-hand side, $\ket{h_-}$, which are the quantum states of the polarization field.  Given an $x,y,z$ coordinate system, we write $\ket{h_+}=\frac{1}{\sqrt{2}}(1,\iu)^{\tran}$ and $\ket{h_-}=\frac{1}{\sqrt{2}}(1,-\iu)^{\tran}$ in the basis $\ket{x}=(1,0)^{\tran},\, \ket{y}=(0,1)^{\tran}$  in the plane perpendicular to $z$. The photon-spin operator along $z$ is then 
\begin{align}
  S_z  = \hbar \left( \ket{h_+}\bra{h_+}-\ket{h_-}\bra{h_-}\right) =\hbar \begin{pmatrix}
      0 & -\iu \\ \iu & 0
  \end{pmatrix}\, ,
\end{align}
with the two eigenstates, $\ket{h_+}, \ket{h_-}$ and  eigenvalues $\pm 1$, which are the helicity values. The general state of the photon is then 
${\displaystyle \ket{\psi} =\alpha_+\ket{h_+}+\alpha_- \ket{h_-}}$ where $1=|\alpha_+|^2+|\alpha_-|^2$. 
The DOFs of the photon spin are captured by  these two coefficients, which produce the probabilities $|\alpha_+|^2, \, |\alpha_-|^2$ for the photon to be in each of the helicities eigenvectors.

\section{Entanglement}
We study entanglement of two fermions and three fermions (triplets) with $s=\nicefrac{1}{2}$ for each fermion. 
\subsection{Two Fermions}
Consider a system with two fermions with spin $s=\nicefrac{1}{2}$, say $A$ and $B$. We focus on their entanglement. Consider  the spin basis, the product of individual spin eigenstates along the $z$ axis,
\begin{align}
    \ket{++}\eqdefA\ket{+}^{A}\ket{+}^{B}, &  \quad \, \ket{+-}\eqdefA\ket{+}^{A}\ket{-}^{B},
    \\
    \ket{-+}\eqdefA\ket{-}^{A}\ket{+}^{B}, & \quad\, \ket{--}\eqdefA\ket{-}^{A}\ket{-}^{B}.
    \label{eq:two-spin-basis-z}
\end{align}

\begin{figure}
 \centering
 a.\includegraphics[scale=0.4]{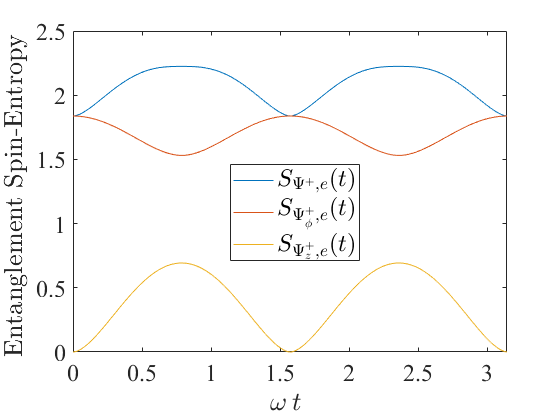}
 \hfil b.\includegraphics[scale=0.4]{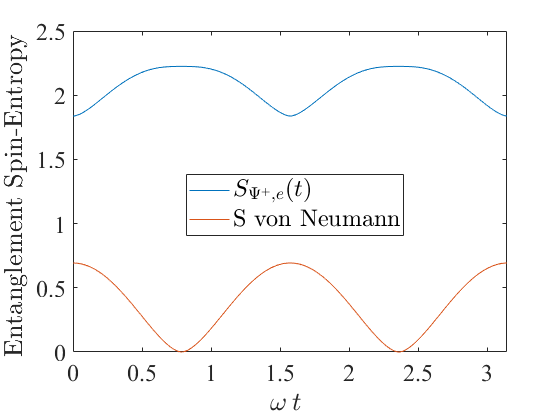}
 \caption{a. Spin-entropy \eqref{eq:spin-entropy-entanglement-half-xi} and its components $\entropyS^{z}(t), \entropyS^{z^{\perp}}(t)$. b. Spin-entropy  vs \vN entropy. The magnitudes are significantly different, and their maxima and minima occur in opposition, at values of  $\omega t$ that are multiples of $\nicefrac{\piu}{4}$.
}
 \label{fig:spin-entropy-GQ-entanglement}
\end{figure}

\begin{theorem}[Two spin $s=\nicefrac{1}{2}$ entanglement]
\label{th:entropy-spin-half-entanglement}
Consider a system with two identical particles, each with $s=\nicefrac{1}{2}$,  in a subspace of spin $0$ along $z$. The Ising-like spin-spin interaction is  described by a  Hamiltonian   $H= \begin{pmatrix} E  & 0   \\ 0 & E \end{pmatrix}$. A spinless environment state $\ket{e}$ with energy $\hbar \gamma$  produces   an orthogonal basis of the system  $\ket{+-}\otimes \ket{e} $ and $\ket{-+}\otimes \ket{e} $. Assume that due to the presence of the environment the two system states  interact via a Hamiltonian  $H^{\mathrm I}=\hbar \begin{pmatrix}  0  &  -\iu \omega     \\ \iu\omega & 0 \end{pmatrix}$ with eigenvalues $\pm \omega$.
Let the initial  state be entangled $\ket{\Psi ^{+},e}={\frac {1}{\sqrt {2}}}(\ket{+-}+\ket{-+})\otimes \ket{e}$. Then, the spin-entropy of  this state
 $ \ket{\Psi ^{+}(t),e(t)}=U(t)  \ket{\Psi^+,e}=\eu^{-\iu \frac{(E+\hbar\gamma)  I+H^{\mathrm I}}{\hbar}\,t}  \ket{\Psi^+,e}$ is
 \begin{align}
 \label{eq:spin-entropy-entanglement-half-xi}
   \entropyS(t)&= -\sin^2 \left(\omega  t\right) \ln \sin^2 \left(\omega  t\right)
 - \cos^2 \left(\omega  t\right) \ln  \cos^2 \left(\omega  t\right)
 \\
 & \quad - \int_0^{2\piu} \frac{(1+\sin (2\omega t)\cos \phi)}{2\piu}  \ln \frac{(1+\sin (2\omega t)\cos \phi) }{2\piu}\, \diff \phi\, ,
\end{align}
which is invariant under $E$ and $\gamma$,   and reaches its maximum  at time $T=\nicefrac{\pi}{4\omega }$, when the state is disentangled  with probability $1$.
\end{theorem}
\begin{proof}
The Hamiltonian  can be diagonalized as
\begin{equation}
  \label{eq:1}
H+\hbar \gamma {\mathrm I}+ H^{\mathrm I}=\frac{\hbar}{2}
   \begin{pmatrix}  1& -\iu  \\ \iu & -1 \end{pmatrix}\begin{pmatrix}  \frac{E}{\hbar} +\gamma+ \omega  & 0  \\ 0 & \frac{E}{\hbar}+\gamma -\omega \end{pmatrix}\begin{pmatrix}  1 & -\iu  \\  \iu & -1   \end{pmatrix}\,.
\end{equation}
 Thus,  the  unitary evolution  operator $U(t)$ is
\begin{align}
   U(t)&=\eu^{-\iu (\frac{E}{\hbar}+\gamma) t} \begin{pmatrix} \cos  \omega t  &\sin  \omega t\\ -\sin  \omega t & \cos  \omega t \end{pmatrix}\,.
   \label{eq:unitary-I}
\end{align}
To evaluate the spin-entropy, we  examine the spin phase space for the two fermions case.
Spin matrices associated with  the two fermions can be written in terms of Pauli matrices $\sigma_x,\sigma_y, \sigma_z$ of the individual fermions as
 \begin{align}
   S_{x,y,z} = \frac{1}{2} \sigma_{x,y,z} \otimes {\mathrm I} + {\mathrm I} \otimes \frac{1}{2} \sigma_{x,y,z}   \quad \text{and} \quad S^2=S_x^2+S_y^2+S_z^2\,,
 \end{align}
written in the basis of products of single particle eigenstates along $z$, namely $\ket{++},\ket{+-},\ket{-+},\ket{--}$. Some of these vectors however are not eigenstates of $S^2$.  A common eigenbasis to $S^2$ and $S_z$, written in terms of the product of single particle eigentstates, is given by  $\big\{\ket{++},  \frac{1}{\sqrt{2}}(\ket{+-}+\ket{-+}), \frac{1}{\sqrt{2}}(\ket{+-}-\ket{-+}),\ket{--} \big\}$. In particular, we are exploring the subspace with spin $0$ along the $z$ axis ($m=0$), i.e.,  the subspace generated by the  two Bell states
\begin{align}
    \ket{\Psi ^{+}}=\ket{\xi_{2,1,0}} ={\frac {1}{\sqrt {2}}}(\ket{+-}+\ket{-+}) \,,\quad
    \ket{\Psi ^{-}}=\ket{\xi_{2,0,0}}={\frac {1}{\sqrt {2}}}(\ket{+-}-\ket{-+})\, ,
    \label{eq:Bell-states}
\end{align}
which are entangled states and eigenstates of $S_z$ with $m=0$. The basis transformation from $\ket{+}\ket{-}, \ket{+}\ket{-} $ to $\ket{\Psi ^{+}},\ket{\Psi ^{-}}$ is  $B=\begin{pmatrix}
\frac{1}{\sqrt{2}} & \frac{1}{\sqrt{2}} \\ \frac{1}{\sqrt{2}} & -\frac{1}{\sqrt{2}}
\end{pmatrix}$ and so, in the basis $\ket{\Psi ^{+},e},\ket{\Psi ^{-},e}$ the unitary evolution operator is
\begin{align}
   U_z(t)&=B U(t) B^{-1}=\eu^{-\iu (\frac{E}{\hbar}+\gamma) t} \begin{pmatrix} \cos  \omega t  &-\sin  \omega t\\ \sin  \omega t & \cos  \omega t\end{pmatrix}\, ,
\end{align}
where the subscript in $U_z$ indicates that this representation is associated with  eigenstates of the spin operators $S_z,S^2$. Thus,
$ \ket{\Psi^+(t),e(t)}=U_z(t)  \ket{\Psi^+,e}=\eu^{-\iu (\frac{E}{\hbar}+\gamma) t} \begin{pmatrix} \cos  \omega , & \sin  \omega t \end{pmatrix}^{\tran}$ with probabilities $\pr(t)=\begin{pmatrix} \cos^2  \omega t,&\sin^2 \omega t \end{pmatrix}^{\tran}$.
We now examine the entropy term of the state $\ket{\Psi^+(t),e(t)}$ in the conjugate basis $\ket{\phi,e}$. From \eqref{eq:xi-phi}
\begin{align}
   \ket{\Psi^+_{z^{\perp}}(t),e(t)}&= \frac{\eu^{-\iu (\frac{E}{\hbar}+\gamma) t} }{\sqrt{2\piu}} \int \left (
   \eu^{\iu \phi} \cos \omega t+  \sin \omega t\right) \ket{\phi,e} \diff \phi\,,
    \end{align}
    implying that
    $
    \rho_{z^{\perp}}(\phi,t)=\nicefrac{1}{2\piu}\big (1+\sin (2\omega t)\cos \phi\big)
$ and the entropy \eqref{eq:spin-entropy-geometric-quantization-multiple-particles} can be written as 
\begin{align}
    \entropyS_{2,\nicefrac{1}{2}}(t)&=\entropyS^{z}_{2,\nicefrac{1}{2}}(t)+\entropyS^{z^{\perp}}_{2,\nicefrac{1}{2}}(t)=-\cos^2  \omega t\ln \cos^2  \omega t-\sin^2  \omega t\ln \sin^2  \omega
    \\
    &\quad  -  \int \frac{1}{2\piu}\big (1+\sin (2\omega t)\cos \phi\big)\ln \frac{1}{2\piu}\big (1+\sin (2\omega t)\cos \phi\big) \, \diff \phi \, ,
\end{align}
and readily evaluated. There is an oscillation of period $\nicefrac{\piu}{2\omega}$. The  entropy is maximized at $T=\nicefrac{\pi}{4\omega }$, where $\ket{\Psi^+(T),e(T)}=\eu^{-\iu (\frac{E}{\hbar}+\gamma) \frac{\pi}{4\omega }}\frac{1}{\sqrt{2}} \begin{pmatrix} 1, & 1 \end{pmatrix}^{\tran}=\eu^{-\iu (\frac{E}{\hbar}+\gamma) \frac{\pi}{4\omega }}\ket{+-,e}$, when with probability 1,  the spin state is  in the disentangled state $\ket{+-,e}$.
\end{proof}
For a visualization of this entropy $ \entropyS_{2,\nicefrac{1}{2}}(t)$ and of its two components, $ \entropyS^{z}_{2,\nicefrac{1}{2}}(t)$ and $ \entropyS^{z^{\perp}}_{2,\nicefrac{1}{2}}(t)$, see Figure~\ref{fig:spin-entropy-GQ-entanglement}a.

We point out that the environment was only modeled as one state, i.e., we did not consider the  pioneer work~\cite{joos1985emergence,zeh1970interpretation} of decoherence.   It would be interesting in future research to  study how the proposed spin-entropy for mixed states can impact the environment on quantum states. 

\subsection{Comparison with Von Neuman Entropy and Werhl Entropy}

Work exists exploring the information content of entangled  physical systems considering  \vN entropy, e.g., \cite{witten2018mini, PhysRevA.53.2046,Calabrese_2004,PhysRevLett.71.666}. In that approach  the required choice of basis functions is  the product of one-particle eigenstates  $\ket{+}\ket{-}, \ket{+}\ket{-} $, so that one can trace out  over one-particle spin eigenstates. Thus, in this basis, the initial state is represented as $\ket{\Psi^+,e}=(\frac{1}{\sqrt{2}},\frac{1}{\sqrt{2}})$  and   its  evolution in this basis is described by
\begin{align}
   \ket{\Psi^+(t),e(t)}=U(t) \ket{\Psi^+,e}=  \eu^{-\iu (\frac{E}{\hbar}+\upGamma) t} \begin{pmatrix} \sin \left(\omega t+\frac{\pi}{4}\right )\\ \cos  \left(\omega t+\frac{\pi}{4}\right ) \end{pmatrix}\,.
\end{align}
After tracing out any of the two particles states,    the probabilities of the resulting mixed state are $\pr_{+}(t)=\sin^2 \left(\omega t+\frac{\pi}{4}\right )$ and $ \pr_{-}(t)=\cos^2 \left(\omega t+\frac{\pi}{4}\right )$ with \vN entropy
\begin{align}
\entropyS^{\text{vN}}(t)=- \cos^2 \left(\omega t+\frac{\pi}{4}\right ) \log\cos^2 \left(\omega t+\frac{\pi}{4}\right )  - \sin^2 \left(\omega t+\frac{\pi}{4}\right )\log\sin^2 \left(\omega t+\frac{\pi}{4}\right )\,,
\label{eq:VonNewmann-entropy}
\end{align}
which is also plotted in Figure~\ref{fig:spin-entropy-GQ-entanglement}b for a visual comparison with the spin-entropy of the original pure state. 

Wehrl's entropy for such mixed state was evaluated by  \cite{Mintert:2004} to conclude that, like \vN entropy,  disentangled states minimize Wehrl entropy. 

Disentanglement occurs at time steps  $ t=\frac{\piu}{4 \omega}+n\frac{\piu}{ \omega};\, n=0,1,\ldots\, $. The spin-entropy is maximized at these times, while the tracing out procedure leads to the minimum of the \vN entropy and the Wehrl's entropy. The  $0$ value for the \vN entropy is  to be expected since this mixed state becomes a pure state at disentanglement. In contrast,  the Bell states (the eigenstates of $S_z, S^2$ for $m=0$)  minimize the spin-entropy and maximize the \vN entropy. These min/max entropy differences are, technically,  due to the spin basis. For the spin-entropy, the spin basis needs to be the eigenfunctions of the two-particles spin operators $S_z, S^2$,  while for the \vN entropy, one must trace out the eigenstates of the individual particle eigenstates. The large entropy magnitude difference everywhere  is due to the spin-entropy's contribution from the $\phi$ variable, the conjugate variable to $z=\cos \theta$.

\subsection{Triplets}

Given a triplet set of spin $\nicefrac{1}{2}$ particles, the total-spin matrices $S_{x,y,z}$ in the basis of products of single particle eigenstates are

\begin{align}
   S_{x,y,z} & = \sigma_{x,y,z} \otimes {\rm I} \otimes {\rm I} + {\rm I} \otimes \sigma_{x,y,z} \otimes {\rm I} +  {\rm I} \otimes {\rm I} \otimes \sigma_{x,y,z}
     \label{eq:spin-z-matrix-three-fermions}
\end{align}
where the Pauli matrices are written in the single particle $z$ axis eigenvectors basis. 
A common basis for both, $S_z$ and $S^2$ is given by  the vectors \cite{Edmonds96}
\begin{align}
     \ket{\xi_{\frac{3}{2},\frac{3}{2}}}&=\ket{+++}, 
     \\ 
     \ket{\xi_{\frac{3}{2},\frac{1}{2}}}&=\frac{1}{\sqrt{3}}\left (\ket{++-}+\ket{+-+}+\ket{-++}\right),\, 
     \\
     \ket{\xi_{\frac{3}{2},-\frac{1}{2}}}&=\frac{1}{\sqrt{3}}\left (\ket{--+}+\ket{-+-}+\ket{+--}\right), \,
     \\
     \ket{\xi_{\frac{3}{2},-\frac{3}{2}}}&=\ket{---},
     \\
     \ket{\xi_{\frac{1}{2},\frac{1}{2}}(\theta_{+})}&=\frac{\cos\theta_{+}}{\sqrt{2}}\ket{+}\otimes(\ket{+-}-\ket{-+})+\frac{\sin\theta_{+}}{\sqrt{2}}(\ket{+-}-\ket{-+})\otimes\ket{+}, 
     \\ 
     \ket{\xi_{\frac{1}{2},-\frac{1}{2}}(\theta_{-})}&=\frac{\cos\theta_{-}}{\sqrt{2}}\ket{-}\otimes(\ket{+-}-\ket{-+})+\frac{\sin\theta_{-}}{\sqrt{2}}(\ket{+-}-\ket{-+})\otimes\ket{-}
     \label{eq:eigenstates-Sz-S2}
 \end{align}
 where the parameters $\theta_{+},\theta_{-}\in [0,2\pi)$ characterize the  degeneracy  of the subspace of spin magnitude $s=1/2$, a four dimensional subspace with only two eigenvalues. 
 
 \subsection{Maximally Entangled States}
According to our spin-entropy, the minimum $\ln 2\piu$ is associated with the eigenstates \eqref{eq:eigenstates-Sz-S2}, i.e.,  entangled states $\ket{\xi_{\frac{3}{2},\frac{1}{2}}},\ket{\xi_{\frac{3}{2},-\frac{1}{2}}}$ and entangled subspaces $\ket{\xi_{\frac{1}{2},\frac{1}{2}}(\theta_{+})},\ket{\xi_{\frac{1}{2},-\frac{1}{2}}(\theta_{-})}$. We propose that the minimum entropy  be the criterion to define {\em maximally entangled states} among all the entangled states. Thus  entangled state W \cite{W2000},  $\ket{\psi_{W}}=\frac{1}{\sqrt{3}}\left (\ket{+--}+\ket{-+-}+\ket{--+}\right)$, is a {\em maximally entangled state}  since it is the  eigenstate $\ket{\xi_{\frac{3}{2},-\frac{1}{2}}}$ and thus has  lowest entropy. 

In contrast, consider the   entangled state GHZ \cite{Greenberger1989}, $\ket{\psi_{GHZ}}=\frac{1}{\sqrt{2}}\left (\ket{+++}+\ket{---}\right)$. Clearly, it is not an eigenstate of either $S^2$ or $S_z$. The entropy will be larger than for all  entangled eigenstates of $S^2, S_z$ and thus it is not a {\em maximally entangled state}. 

Ranking  entangled states according to the spin-entropy provides an information content evaluation that may be helpful when devising quantum physical processes. The lower the entropy, the further away from decoherence. 

\section{Conclusions}
\label{sec:conclusion}
The concept of spin-entropy in a spin phase space  is proposed. The spin phase space of a particle is defined via the already existing Geometric Quantization method that  quantize  a sphere surface.  The  operators associated with the spherical polar representation do not commute, yielding the uncertainty principle for the spin values in phase space.  The states in spin phase space  are the simultaneous projections of a spin state onto the  $z$ axis eigenstates  and onto the polarization angle sates, which generates the plane perpedicular to the $z$ axis.   The proposed spin-entropy captures the randomness present in the spin state for a specified $z$ axis. The $z$ axis for a system of fermions is  defined as the direction of the magnetic field present in the system, and it can vary over time.   In the case of a photon, the $z$ axis is the direction of the propagation, where the helicity is defined.  The formulation is general for a system of many spin particles and extends to quantum fields. We studied spin-entropy for single  particles  with spin $\nicefrac{1}{2} $, spin $1$, photons,  and for  two and three entangled fermions of spin $\nicefrac{1}{2}$.

We have examined  entangled states and their time evolution.  Bell's entangled states that are eigenstates of the spin $S_z, S^2$ operators with total zero spin value along the $z$ axis  have lower spin-entropy than that of the product of one particle states (disentangled states).
In contrast, the \vN entropy and Wehrl entropy are maximized at Bell's entangled states and minimized at disentangled states. 
We  studied the dynamics of entangled states with an Ising-like Hamiltonian model of the interaction between spins, and  a model of the environment as one state and its impact on a Hamiltonian mediating the interaction between Bell's entangled states. In a simulation, the time evolution of such Bell entangled states the entropy  increases up to its maximum value, and disentanglement then occurs. 

We then analyzed some quantum states of three fermions of spin $\nicefrac{1}{2}$ and suggested that maximum entangled states should be defined by the entropy value. The lower the entropy of an entangled state,  the larger the entanglement.

\section{Acknowledgement} This paper is partially based upon work supported by both the National Science Foundation under Grant No.~DMS-1439786 and the Simons Foundation Institute Grant Award ID 507536 while the first author was in residence at the Institute for Computational and Experimental Research in Mathematics in Providence, RI, during the spring 2019 semester ``Computer Vision''  program. We thank Dries Sels for various conversations and bringing to our attention the work  on coherent spin states.

\bibliographystyle{abbrv}
\bibliography{gk01}

\end{document}